# Topological Design of Minimum Cost Survivable Computer Communication Networks: Bipartite Graph Method

Kamalesh V.N
Research Scholar, Department of Computer Science and Engineering,
Sathyabama University,
Chennai, India
Kamalesh_v_n@yahoo.com

S K Srivatsa
Senior Professor, Department of Instrumentation and Electronics Engineering,
St. Joseph College of Engineering,
Chennai, India.
profsks@rediffmail.com

*Abstract*- A good computer network is hard to disrupt. It is desired that the computer communication network remains connected even when some of the links or nodes fail. Since the communication links are expensive, one wants to achieve these goals with fewer links. The computer communication network is fault tolerant if it has alternative paths between vertices, the more disjoint paths, the better is the survivability. This paper presents a method for generating k-connected computer communication network with optimal number of links using bipartite graph concept.

*Keywords ; computer network, link deficit algorithm, wireless network, k-connected networks, survivable network, bipartite graph.*

## I INTRODUCTION

The topological design of a network assigns the links and link capacities for connecting the network nodes. This is a critical phase of network synthesis, partly because the routing, flow control and other behavioral design algorithms rest largely on the given network topology. The topological design has also several performance and economic implications. The node locations, link connections and link speeds directly determine the transit time through the network. For reliability or security considerations, some networks may be required to provide more than one distinct path for each node pair, thereby resulting in a minimum degree of connectivity between the nodes [10]. The performance of a fault-tolerant system should include two aspects, computational efficiency and network reliability. When a component or link fails, its duties must be taken over by other fault free components or links of the system. The network must continue to work in case of node failure or edge failure. Different notions of fault tolerance exist, the simplest one corresponding to connectivity of the network, that is, the minimum number of nodes which must be deleted in order to destroy all paths between a pair of nodes. The maximum connectivity is desirable since it corresponds to not only the maximum fault tolerance of the network but also the maximum number of internally disjoint paths between any two distinct vertices. However, connectivity number can be at most equal to the degree of the network graph [7, 8]. The goal of the topological design of a computer communication network is to achieve a specified performance at a minimal cost [5]. Unfortunately, the problem is completely intractable [1]. If the network under consideration has n distinct nodes and p distinct possible bandwidth then the size of the space of potential topologies would be $(p+1)^{n(n-1)/2}$ for the values n=10 and p=3 the size of the search space would be $1.2 \times 10^{27}$. The fastest available computers cannot optimize a 25 node network, let alone a 100 node network. A reasonable approach is to generate a potential network topology (starting network) and see if it satisfies the connectivity and delay constraints. If not, the starting network topology is subjected to a small modification ("perturbation") yielding a slightly different network, which is now checked to see if it is better. If a better network is found, it is used as the base for more perturbations. If the network resulting from perturbation is not better, the original network is perturbed in some other way. This process is repeated till the computer budget is used up. [2, 3, 5].

A fundamental problem in network design is to construct a minimum cost network that satisfies some specified connectivity requirements between pair of nodes. One of the most well suited problems in this framework is the survivable network design problem, where we are given a computer communication network with costs on its edges, and a connectivity requirement $r_{ij}$ for each pair i,j of nodes. The goal is to find a minimum cost subset of edges that ensures that there exist $r_{ij}$ disjoint paths for each pair i,j of nodes. Where all $r_{ij} \in \{0,k\}$, for some integer k, we will refer to these problems as k-connectivity of a computer communication network and k-vertex connectivity respectively [13].

Because of the importance of the problem, few methods for generating k-connected networks are proposed in the literature. In the method due to Steigletiz, et. al.[4], the heuristic begins by numbering the nodes at random. This randomization lets the heuristic to generate many topologies from the same input data. Further this method involves repeated searching of nodes when conflicts occur. This demands more computational effort. In the method [9] the nodes are numbered arbitrarily. The decimal number of each node is converted into a k bit Gray code. Thus each node has a Gray code associated with it. There exist a link between any two nodes whose Gray codes differ only in one place are connected. Thus every node gets connected to k nodes and has a degree of k. However the limitation of this method is again arbitrarily numbering of nodes and this method is applicable only when number of nodes in the network is $2^k$. In the method [11] the nodes are numbered arbitrarily and assume that the nodes of the network are equispaced and lie on a circle, that is their method is applicable only when the nodes of the network form a regular polygon.

In our earlier method [12] for generating k-connected survivable network topology, the geographical positions of the nodes are given. To start with, the nodes are labeled using some symbols. Given the cost of establishing link between pair of nodes the same is represented in the form of the matrix. The accumulated cost for every node is computed. The accumulated cost is sorted in the increasing order. The index value of the sorted list is assigned as representative number for the nodes. Links are established between nodes. The details can be found in [12]. However in this method redundant links are more in number.

In view of overcoming the above said limitations, in this paper we proposed a novel technique, an approach based on bipartite graph which ensures generation of minimum cost k-connected survivable network topologies.

The section 2 explains the proposed method. Section 3 give the illustration of proposed method finally the paper concludes in section 4.

## II PROPOSED METHOD

This section presents the proposed technique for generating k-connected survivable network topologies. The geographical positions of the nodes are given. To start with, the nodes are labeled using some symbols. Given the cost of establishing link between pair of nodes the same is represented in the form of the matrix. The accumulated cost for every node is computed. Sort the accumulated costs in the increasing order. Assign the index value of the sorted list as representative number for the nodes. Partition the node set of the network into two sets $V_1$ and $V_2$ such that $|V_1| = k$ and $|V_2| = n-k$ i.e., $V_1 = \{1, 2, 3, 4,…k\}$ and $V_2 = \{k+1, k+2,…n\}$, where $n > k$. Construct a bipartite graph $G(V_1 : V_2, E)$ with nodes $V_1$ and $V_2$. The network graph so obtained is k-connected network graph.

*Algorithm*: Generate minimum cost k-connected survivable network topology.

Input:
    (i) n- nodes of the network
    (ii) Cost associated with each pair of nodes
    (iii) k- connectivity number (k < n)

Output:
    Minimum cost k-connected survivable network topology.

Method:
1. Geographical positions of the n distinct nodes are given.
2. Name the nodes using some symbols.
3. Construct a cost matrix using cost associated with each pair of nodes.
4. Compute the accumulated cost for every node.
5. Sort the result of step 4 using appropriate sorting technique in increasing order.
6. Assign the index value of the sorted list as the representative number for the nodes.
7. Partition the node set V into two sets $V_1$ and $V_2$ such that $|V_1| = k$ and $|V_2| = n-k$ and $V_1 = \{1, 2, 3, 4,…k\}$ and $V_2 = \{k+1, k+2,…n\}$.
8. Construct a complete bipartite graph with node sets $V_1$ and $V_2$. ie for i=1 to k
      for j=k+1 to n
        establish link between i & j.
      end of for
    end of for
The network graph obtain is a k-connected network graph.

Algorithm end

This method is link optimal compared to the methods [12]. The number of links required for the generation of the k-connected network with n nodes in this method is k(n-k). Where as in the method [12], the number of links required for the generation of the k-connected network with n nodes is
(n-1)+(n-2)+…+(n-k).
Further (n-1)+(n-2)+…+(n-k) > (n-k)+(n-k)+…+(n-k) for k >1.
ie (n-1)+(n-2)+…+(n-k) > k (n-k).
Hence the number of links used in method [12] is greater than the number of links used in the proposed method.
This method is also link optimal compared to the methods [4,11], for k>n/2. In the methods [4,11], the number of links required for the generation of the k-connected network with n nodes is nk/2, and k(n-k) > kn/2 for all k > n/2.

The comparative analysis for link optimization of various methods are tabulated in table 1

TABLE-I COMPARATIVEI ANALYSIS FOR LINK OPTIMIZATION

| Method | Number of links used to generate k-connected network with n nodes | Comparative analysis |
|---|---|---|
| The method presented in this paper. | k(n-k) | |
| Kamalesh V.N et al.[12] | (n-1) + (n-2 ) + ... + (n-k) | k(n-k) < (n-1)+(n-2)+... + (n-k), for all n and k. Hence, the method presented in this paper is optimal compared due to Kamalesh V.N et al. |
| S. Latha. et al. [11] | kn/2 | k(n-k) < kn/2, for all k> n/2. Hence, the method presented in this paper is optimal compared to method due to S. Latha et al. [11], for all k > n/2. |
| K. Steiglitz . et al. [4] | kn/2 | k(n-k) < kn/2, for all k> n/2. Hence, the method presented in this paper is optimal compared to method due to K. Steiglitz et al. [4], for all k > n/2. |

## III RESULTS ILLUSTRATION

In this section we illustrate the proposed method for 7 distinct nodes. The network connectivity k is assumed to be 3.
Let us consider the 7 nodes with some labels as shown in Fig-1.

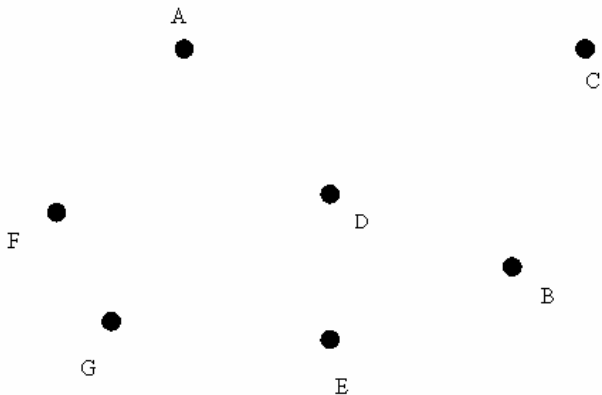

Figure1:– Name the nodes of the given Network with some symbols

The Table-2 gives the cost matrix constructed out of the cost associated with pair of nodes. The last column of Table-1 gives the accumulated cost for every node.

TABLE-II COST ASSOCIATED WITH EVERY PAIR OF NODES

|   | A | B | C | D | E | F | G | Accumulate Cost |
|---|---|---|---|---|---|---|---|---|
| A | 0 | 4 | 2 | 4 | 3 | 1 | 5 | 19 |
| B | 4 | 0 | 4 | 5 | 2 | 3 | 2 | 20 |
| C | 2 | 4 | 0 | 1 | 4 | 3 | 1 | 15 |
| D | 4 | 5 | 1 | 0 | 2 | 2 | 4 | 18 |
| E | 3 | 2 | 4 | 2 | 0 | 1 | 10 | 22 |
| F | 1 | 3 | 3 | 2 | 1 | 0 | 3 | 13 |
| G | 5 | 2 | 1 | 4 | 10 | 3 | 0 | 25 |

Table-3 gives the numbers associated for every node based on the sorted accumulated cost.

TABLE-III: NODE NUMBERING

| Nodes | Accumulated Cost | Nodes Numbers |
|---|---|---|
| A | 19 | 4 |
| B | 20 | 5 |
| C | 15 | 2 |
| D | 18 | 3 |
| E | 22 | 6 |
| F | 13 | 1 |
| G | 25 | 7 |

Fig-2 shows the nodes labeled using the sorted accumulated cost. Whereas, fig-3 shows the minimum cost 1-connected network topology.

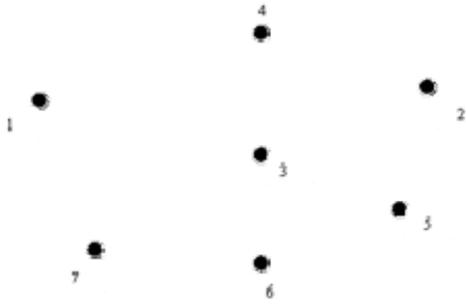

Figure2:- Nodes are numbered using cost matrix

Since k=3 therefore $|V_1| = 3$ and $|V_2| = 7-3 = 4$

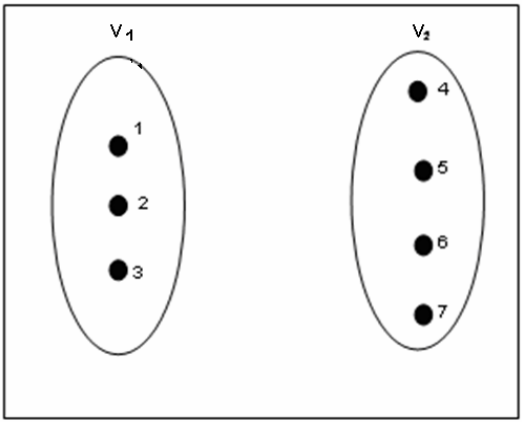

Figure3:– The node set is partitioned into two sets $V_1$ and $V_2$

Construct the bipartite graph G ($V_1$: $V_2$, E)

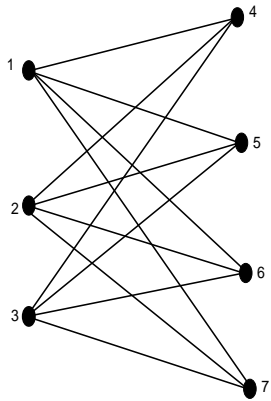

Figure4:-Complete Bipartite Graph

The resultant 3-connected network graph is shown in Figure5.

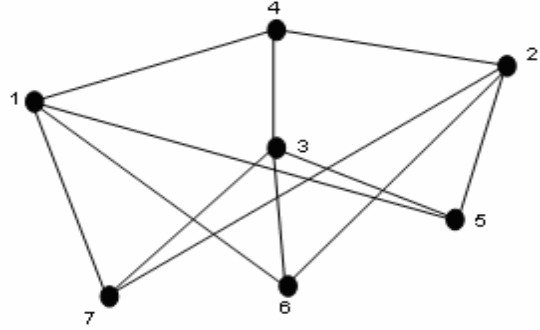

Figure5: 3-connected network graph

## IV CONCLUSIONS

The topological design of a network assigns the links and link capacities for connecting the network nodes. This is a critical phase of network synthesis, partly because the routing, flow control and other behavioral design algorithms rest largely on the given network topology. The goal of the topological design of a computer communication network is to achieve a specified performance at a minimal cost. In this paper we presented a generic method to generate a minimum cost k-connected survivable network topology. The main strength of the proposed method is that it is very simple and straightforward. Also, unlike other existing methods, the proposed method does not make any specific assumption to generate a network. Further in this method the number of links used is optimal compared to [11,12].

## VI BIOGRAPHY

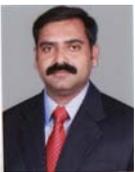

Prof. Kamalesh V. N received the Bachelor of Science degree from University of Mysore, India. Subsequently, he received Master of Science in Mathematics degree and Master of Technology in Computer Science & Technology degree from University of Mysore, India. He secured 14$^{th}$ rank in Bachelor of Science and 4$^{th}$ rank in Master of Science from University of Mysore. Further, he was National Merit Scholar and subject Scholar in Mathematics. He is working as Head, Department of Computer Science & Engineering at JSS Academy of Technical Education, Bangalore, affiliated to Visvesvaraya Technological University, Belgaum, Karnataka, India. He has taught around fifteen different courses at both undergraduate and post graduate level in mathematics and Computer science and engineering. His current research activities pertain to Computer networks, Design and Analysis of algorithms, Graph theory and Combinatorics, Finite Automata and Formal Languages. His paper entitled "On the assignment of node number in a computer communication network" was awarded certificate of merit at World Congress on Engineering and Computer Science 2008 organized by International Association of Engineers at UC Berkeley, San Francisco, USA. He is currently research candidate pursuing Ph.D degree from Sathyabama University, Chennai, India.

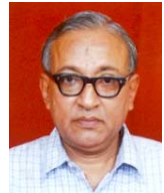

Dr. S.K. Srivatsa received the Bachelor of Electronics and Telecommunication Engineering degree from Jadavpur University, Calcutta, India. Master's degree in Electrical Communication Engineering and Ph.D from the Indian Institute of Science, Bangalore, India. He was a Professor of Electronics Engineering in Anna University, Chennai, India. He was a Research Associate at Indian Institute of Science. Presently he is working as senior Professor in department of Instrumentation & Control Engineering at St. Joseph College of Engineering, Chennai, India. He has taught around twenty different courses at undergraduate and post graduate level. His current research activities pertain to computer networks, Design and Analysis of algorithms, coding Theory and Artificial Intelligence & Robotics.